\documentstyle[prl,aps,preprint]{revtex}

\begin{document}
\draft

\title{Thermalization of a Brownian particle via
coupling to low-dimensional chaos}
\author{C. Jarzynski}
\address{Institute for Nuclear Theory, University of
Washington,\\
Seattle, WA~~98195}
\date{\today}
\maketitle

\begin{abstract}
It is shown that a paradigm of classical
statistical mechanics --- the thermalization of a
Brownian particle --- has a
low-dimensional, deterministic analogue:
when a heavy, slow system is coupled to fast
deterministic chaos, the resultant forces
drive the slow degrees of freedom toward a state of
statistical equilibrium with the fast degrees.
This illustrates how concepts useful in
statistical mechanics may apply in situations
where low-dimensional chaos exists.
\end{abstract}


Since the study of chaotic dynamics has
clarified fundamental issues in classical
statistical mechanics \cite{rasetti}, it is worthwhile
to consider the converse:
when does intuition from statistical mechanics
carry over to {\it low-dimensional} chaos?
We all know, for instance, that a heavy particle
immersed in a heat bath ---a Brownian particle ---
is subject to both an average frictional force,
and stochastic fluctuations around this average,
and that the balance between these two
{\it thermalizes} the particle.
Now suppose the ``Brownian'' particle is coupled to a fast,
low-dimensional, chaotic trajectory,
rather than to a true heat bath.
It is known that the particle then feels a
dissipative force\cite{ott,wilk,br}; does the particle also
(in some sense yet to be defined)
``thermalize'' with the chaotic trajectory?
That is, does the fast chaos behave
as a kind of ``miniature heat reservoir'',
exchanging energy
with the particle in a way that brings the two into
statistical equilibrium?
In this paper, we pursue this question by considering the
reaction forces acting on a heavy, slow system (our Brownian
particle) due to its coupling to a light, fast trajectory.
When the fast motion
is chaotic, the forces on the particle include
a conservative force, and two velocity-dependent
forces, one magnetic-like, the other dissipative\cite{br}.
However (as in the case of coupling to a true thermal
bath), there also exists a rapidly fluctuating, effectively
{\it stochastic} force, which has not been studied
in detail.
We describe an approach which incorporates
this force, with the others, into a unified
framework.
It is shown that the inclusion of this stochastic
force --- related to the frictional force
by a fluctuation-dissipation relation\cite{br} ---
causes the slow Brownian particle and
the fast chaotic trajectory to evolve toward
statistical equilibrium.

This result provides some justification for
applying statistical arguments (involving, e.g., relaxation
toward equipartition of energy) to physical situations
of only a few degrees of freedom.
A discussion of examples --- including one-body
dissipation in nuclear dynamics\cite{wf},
the Fermi mechanism of cosmic ray acceleration\cite{fermi},
and the diffusive transport of comets\cite{comets} ---
where such ``thermal'' arguments may
provide insight into the physics behind more explicit
calculations, will be presented in Ref.\cite{long}.

As a starting point for our discussion, we consider
the framework of Ref.\cite{br}, where
the position {\bf R} of the slow particle parametrizes
the Hamiltonian $h$ governing the fast motion:
$h=h({\bf z},{\bf R})$, where {\bf z} denotes the
fast phase space coordinates.
(The nature of the fast system will remain unspecified, but
we take it to have {\it a few}, $N\sim 2$, degrees of freedom.)
This classical version of the
Born-Oppenheimer framework
has received considerable interest in recent
years\cite{wilk,br,recent}.
We assume that, if {\bf R} were held {\it fixed},
then a fast trajectory evolving under $h$
would ergodically and chaotically explore
its {\it energy shell} (surface of constant $h$)
in the fast phase space.
This sets a fast time scale, $\tau_{f}$,
which we may take to be the Lyapunov time
associated with the fast chaos.
A slow time scale, $\tau_{s}$, is set by
the motion of the slow particle:
it is the time required for the
Hamiltonian $h$ to change significantly.
We assume $\tau_f\ll\tau_s$;
thus, the fast trajectory ${\bf z}(t)$
evolves under a slowly time-dependent Hamiltonian $h$.
The full Hamiltonian for the combined system
of slow and fast degrees is given by
$H({\bf R},{\bf P},{\bf z})=
P^2/2M+h({\bf z},{\bf R})$,
where ${\bf P}$ is the momentum of the slow
particle, and $M$ is its mass.
$({\bf R},{\bf P},{\bf z})$ thus specifies
a point in the full phase space of slow and
fast variables.
It is assumed that surfaces of constant $H$
are bounded in the full phase space.

Given this formulation, the force
on the slow particle is
${\bf F}(t)=-\partial h/\partial{\bf R}$,
evaluated along the trajectory ${\bf z}(t)$.
{}From the point of view of the slow particle,
this force fluctuates rapidly,
so it is natural to separate
${\bf F}(t)$ into a
slowly-changing {\it average} component,
and rapid fluctuations
$\tilde{\bf F}(t)$ around this average.
In Ref.\cite{br}, Berry and Robbins introduce
an approximation scheme for obtaining
the net {\it average} reaction force.
At leading (zeroth) order of approximation, the
{\it ergodic adiabatic invariant}\cite{ott}
dictates the energy of the fast system as a
function of the slow coordinates, and this
energy in turn serves as a potential for the slow
system, giving rise to a conservative
``Born-Oppenheimer'' force ${\bf F}_0$.
At next order, the Berry-Robbins framework yields
two velocity-dependent reaction forces:
{\it deterministic friction} (${\bf F}_{df}$) and
{\it geometric magnetism} (${\bf F}_{gm}$)\cite{memory}.
Geometric magnetism
is a gauge force related to the geometric phase;
deterministic friction (see also Ref.\cite{wilk})
describes the irreversible flow of energy from the slow
to the fast variables.
Thus, while at leading order the fast
degrees of freedom create a potential well
for the slow degrees, at first order the fast motion
effectively adds a magnetic field, and drains
the slow system of its energy.

What about the effets of the rapidly fluctuating
component, $\tilde{\bf F}(t)$?
If the analogy with ordinary
Brownian motion is correct and some sort of
statistical equilibration occurs, then $\tilde{\bf F}(t)$
ought to play a central role in the process.
We now describe a framework which
incorporates the effects of
$\tilde{\bf F}(t)$ into a description of the slow particle's
evolution.

In our framework we consider
an {\it ensemble} of systems.
Each member of the ensemble consists of a single slow
particle coupled to a single fast
trajectory, and represents one possible realization
of the combined system of slow and fast variables.
Representing this ensemble by a density
$\phi$ in the full
phase space, Liouville's equation is:
\begin{equation}
\label{eq:liou}
{\partial\phi\over\partial t}+
{{\bf P}\over M}\cdot{\partial\phi\over\partial{\bf R}}-
{\partial h\over\partial{\bf R}}\cdot
{\partial\phi\over\partial{\bf P}}+
\{\phi,h\}=0,
\end{equation}
where $\{\cdot,\cdot\}$ denotes the Poisson bracket
with respect to the fast variables, {\bf z}.
Henceforth, we will ignore all information about the
fast trajectory except its energy,
$E(t)\equiv h[{\bf z}(t),{\bf R}(t)]$
(which evolves on the slow time scale\cite{ott}).
Thus, what we are really after is the evolution of
$W({\bf R},{\bf P},E,t)$, the distribution of our
ensemble in the reduced space where all fast variables
other than $E$ have been projected out.
In this reduced space, $\tilde{\bf F}(t)$ is {\it stochastic},
which in turn suggests that $W$ evolves {\it diffusively}.

The derivation of an evolution equation for $W$ is somewhat
involved, and is sketched in the Appendix.
Here we simply state the result:
\begin{equation}
\label{eq:central}
{\partial W\over\partial t}=
-{{\bf P}\over M}\cdot{\partial W\over\partial{\bf R}}+
\hat{\bf D}\cdot({\bf u}W)+
{\epsilon\over 2}
\hat D_i\Biggl[\Sigma L_{ij}\hat D_j
\Biggl({W\over\Sigma}\Biggr)\Biggr].
\end{equation}
(Summation over repeated indices is implied.)
To explain notation, we first define
$\Omega(E,{\bf R})=\int d{\bf z}\,
\theta[E-h({\bf z},{\bf R})]$.
Then $\Sigma\equiv\partial\Omega/\partial E$,
and ${\bf u}\equiv-(1/\Sigma)
(\partial\Omega/\partial{\bf R})$.
Next,
\begin{equation}
\hat{\bf D}\equiv
{\partial\over\partial{\bf P}}-
{{\bf P}\over M}{\partial\over\partial E}.
\end{equation}
$L_{ij}(E,{\bf R})$ is an integrated correlation
function defined by Eq.\ref{eq:lij}.
Finally, $\epsilon\sim\tau_{f}/\tau_{s}\ll 1$
is an ordering parameter; Eq.\ref{eq:central} is valid
to ${\it O}(\epsilon)$.

$\Omega$, $\Sigma$, and {\bf u}
have simple interpretations in terms of the {\it energy
shell} $(E,{\bf R})$ [the surface
{\it in {\bf z}-space} defined by $h({\bf z},{\bf R})=E$].
$\Omega(E,{\bf R})$ is the volume of {\bf z}-space
enclosed by this shell.
$\Sigma(E,{\bf R})=\int d{\bf z}\,\delta(E-h)$
represents the statistical weight of the shell
--- i.e.\ the amount of fast phase space occupied by
this shell ---
and is useful for evaluating energy
shell averages:
\begin{equation}
\langle Q\rangle_{E,{\bf R}}=
{1\over\Sigma(E,{\bf R})}
\int d{\bf z}\,\delta(E-h)\,Q({\bf z}),
\end{equation}
where $\langle Q\rangle_{E,{\bf R}}$
denotes the average of $Q({\bf z})$ over the
energy shell $(E,{\bf R})$.
Finally, ${\bf u}(E,{\bf R})=\langle
\partial h/\partial{\bf R}\rangle_{E,{\bf R}}$.

What does Eq.\ref{eq:central} reveal about
the reaction forces on the slow particle?
Below, we outline calculations
behind the following assertions regarding
the content of Eq.\ref{eq:central}:
(1) it reproduces the average reaction
forces ${\bf F}_0$, ${\bf F}_{df}$, and ${\bf F}_{gm}$;
(2) it describes the effects of the rapidly fluctuating
force $\tilde{\bf F}(t)$;
and (3) it predicts that the Brownian particle does
indeed thermalize with the fast trajectory.
For a more detailed treatment of this problem,
see Ref.\cite{long}.

First, letting ${\cal E}=P^2/2M+E$ denote the total
energy of the system, note that $\hat{\bf D}{\cal E}=0$.
Thus, $\hat{\bf D}$ is a constrained derivative:
$\hat{\bf D}=(\partial/\partial{\bf P})_{\cal E}$,
where the notation indicates that ${\cal E}$, not $E$, is
held fixed.
This means that the evolution depicted by Eq.\ref{eq:central}
takes place along surfaces of constant ${\cal E}$ in
$({\bf R},{\bf P},E)$-space;
this is simply a statement of energy conservation.

Next, if we explicitly separate drift terms from
diffusion terms, Eq.\ref{eq:central} becomes
\begin{equation}
\label{eq:driftdiff}
{\partial W\over\partial t}=
-{\partial\over\partial{\bf R}}\cdot\Biggl(
{{\bf P}\over M}W\Biggr)
-{\partial\over\partial{\bf P}}\cdot\Biggl(
{\bf f}W\Biggr)
+{\epsilon\over 2}
{\partial^2\over\partial P_i\partial P_j}
\Biggl(L_{ij}W\Biggr).
\end{equation}
Here, the derivatives w.r.t.\ ${\bf P}$ are
the constrained derivatives
$(\partial/\partial{\bf P})_{\cal E}$, and
\begin{eqnarray}
\label{eq:f}
{\bf f}=-{\bf u}(E,{\bf R})-\epsilon
K\cdot{{\bf P}\over M},\\
\label{eq:fluctdiss}
K_{ij}(E,{\bf R})={1\over 2\Sigma}{\partial\over\partial E}
\Biggl(\Sigma L_{ij}\Biggr).
\end{eqnarray}
In Eq.\ref{eq:driftdiff}, ${\bf f}$
plays the role of a {\it drift coefficient} for the slow
momentum,
and thus represents the {\it average force
acting on the Brownian particle}.
A comparison with
Ref.\cite{br}, Sec.\ 2, reveals that the
first term of {\bf f} is the leading
(Born-Oppenheimer) force ${\bf F}_0$;
the second is a sum of the
two velocity-dependent forces,
${\bf F}_{df}$ and ${\bf F}_{gm}$:
if we express the matrix $K$ in Eq.\ref{eq:f}
as the sum of its symmetric and
anti-symmetric components, then the former gives us
${\bf F}_{df}$, the latter ${\bf F}_{gm}$.
Eq.\ref{eq:driftdiff} thus reproduces the
{\it average} forces
${\bf F}_0$, ${\bf F}_{df}$, and ${\bf F}_{gm}$
acting on the Brownian particle.

The last term in Eq.\ref{eq:driftdiff} describes the
{\it diffusion of slow momenta} due to the
fluctuating force $\tilde{\bf F}(t)$.
The diffusion coefficient is the matrix $L$, or more
precisely its symmetric component $L^{sym}$.
By Eq.\ref{eq:fluctdiss}, however, $L^{sym}$ is related
to $K^{sym}$, which as mentioned is responsible for
the {\it dissipative} force acting on the slow particle.
Eq.\ref{eq:fluctdiss}
thus emerges as a fluctuation-dissipation relation.
This relation was first noted by Berry and Robbins\cite{br}.

Finally, do the
forces acting on the slow Brownian particle cause it to
thermalize with the fast trajectory?
To answer, we must first define what we mean by
``thermalization'' in the context of the present
problem (where temperature plays no role).
In ordinary statistical mechanics, {\it thermalization}
means, fundamentally, a {\it statistical sharing of
the total energy} $X$:
after a Brownian particle has long been in contact with
a heat bath, the probability for finding it in some
state of energy $x$ is simply proportional to the amount of
phase space available for the bath to have the remaining
energy, $X-x$.
(This leads to the Boltzmann factor $P\propto\exp -x/k_BT$.)
Similarly in the present context, we
take the ``thermalization'' of slow and fast degrees to mean a
statistical sharing of the total energy $\cal E$:
the slow and fast variables have {\it thermalized},
if the probability for finding the former in a state
$({\bf R},{\bf P})$ is simply proportional to the amount
of phase space available for the latter to have energy
$E={\cal E}-P^2/2M$, namely $\Sigma(E,{\bf R})$.
(Thus, $\Sigma(E,{\bf R})$ plays the role of the
Boltzmann constant here.)
For our {\it ensemble}, this condition implies that an
initial distribution $W({\bf R},{\bf P},E,t_0)$
evolves toward one that has the form
\begin{equation}
\label{eq:asymp}
W({\bf R},{\bf P},E,t_\infty)=G({\cal E})\,\Sigma(E,{\bf R}).
\end{equation}
[$G({\cal E})$
is determined by the distribution of total energies,
$\eta({\cal E})$, which remains constant.]

We now make some formal arguments
to show that Eq.\ref{eq:asymp} indeed represents the
ultimate fate of a distribution $W$ evolving under
Eq.\ref{eq:central}.
Consider an {\it entropy}
$S[W]\equiv-\int W\ln (W/\Sigma)$,
where $\int\equiv\int d^3R\int d^3P\int dE$.
Using the identity
$\partial\Sigma/\partial{\bf R}=-
(\partial/\partial E)(\Sigma{\bf u})$,
Eq.\ref{eq:central} gives
\begin{equation}
\label{eq:dsdt}
{dS\over dt}=
{\epsilon\over 2}\int
{\Sigma^2\over W}
L_{ij}\Gamma_i\Gamma_j\ge 0,
\end{equation}
where ${\bf\Gamma}=\hat {\bf D}(W/\Sigma)$.
(The inequality follows from the fact that the
eigenvalues of $L^{sym}$ are non-negative.
A proof of the latter is given in Ref.\cite{long};
less formally, recall that the eigenvalues of
$L^{sym}$ are diffusion coefficients, and as such
have no business being negative.)
Now, the distribution of total energies,
$\eta({\cal E})$,
is conserved as $W$ evolves with time.
However, within the set of all densities $W$
corresponding to a particular $\eta({\cal E})$,
$S[W]$ is bounded from above\cite{long}.
Thus as $W$ evolves with time, the value of $S[W]$
never exceeds a certain upper limit.
Since $dS/dt\ge 0$, the entropy
must eventually {\it saturate},
i.e.\ ${\bf\Gamma}\rightarrow 0$ as
$t\rightarrow\infty$\cite{nofrict}.
This in turn implies that
\begin{equation}
\label{eq:thermal}
W({\bf R},{\bf P},E,t)\rightarrow
g({\cal E},{\bf R},t)\,\Sigma(E,{\bf R}).
\end{equation}
However,
Eq.\ref{eq:thermal} is a solution of Eq.\ref{eq:central}
only if $g$ is independent of both {\bf R} and $t$,
so we finally conclude that
$W\rightarrow G({\cal E})\,\Sigma(E,{\bf R})$
asymptotically with time.
Thus, the ensemble {\it thermalizes}, in the sense
defined in the previous paragraph;
this is our central result.

This result may be restated as follows\cite{long}.
If we start with a fast chaotic, ergodic system,
which we then enlarge
by coupling a few slow degrees of freedom
to the fast ones, then the combined system is itself
ergodic (over the surface of constant $H$)
in the enlarged phase space.
Thus the property of ergodicity is promoted, from the
fast phase space, to the
full phase space of slow and fast variables.

Note also that this ``thermalization''
proceeds on a time scale much longer
than that characterizing the chaotic evolution
($\tau_{f}$).
This is again similar to the case of ordinary Brownian
motion --- where such a separation of time scales
is central\cite{kubo} --- but stands
in contrast to the more familiar examples
of low-dimensional chaos (e.g.\ the $N=2$ Sinai
billiard\cite{sinai}), where the {\it mixing time}
and the Lyapunov time are comparable.

It is no new thing to say that a chaotic, ergodic
trajectory offers a low-dimensional
($N\sim 2$) analogue for a truly thermal ($N\gg 1$)
system.
The novelty of the present work is that it extends
this analogy to encompass the important paradigm of
Brownian motion, where the thermal system or chaotic
trajectory is coupled to a few degrees of freedom
characterized by a much longer time scale.
Then, in either case, the forces acting on the slow
system drive it toward
a state of genuine {\it statistical} equilibrium
with its environment.

Finally, it would be interesting to study
the quantal version of this problem.
Srednicki\cite{sred} has recently argued that
concepts from quantum chaos may provide a solid
foundation for quantum statistical mechanics.
The focus in Ref.\cite{sred} is on genuinely thermal
systems ($N\gg 1$), and does not
deal specifically with the case when a few degrees
of freedom are slower than the rest.
Nevertheless, Srednicki's application of {\it Berry's
conjecture}\cite{bc} to the quantal evolution
of a classically chaotic system
might serve as a guide to a quantal analysis
of the purely classical problem studied here.
(To the best of my knowledge, no one has looked explicitly at
the application of Berry's conjecture to a system
which classically exhibits two widely separated time
scales.)

It is a pleasure to acknowledge that conversations
with Greg Flynn, Allan Kaufman, Robert Littlejohn,
Jim Morehead, and W\l adek \' Swi\c atecki were
very useful in obtaining the results presented in
this papers.
This work has been supported by the Department
of Energy under Grants No.\
DE-AC03-76SF00098 and DE-FG06-90ER40561,
and by the National Science Foundation under
Grant No.\ NSF-PYI-84-51276.

{\bf Appendix}

Here we sketch the derivation of Eq.\ref{eq:central}
from Eq.\ref{eq:liou}, using what is
essentially (though not explicitly) a
multiple-time-scale analysis,
and is similar to that of Ref.\cite{br}.

To begin, we use our adiabaticity
parameter $\epsilon\ll 1$ to formally incorporate
into Eq.\ref{eq:liou} the assumption that
$({\bf R},{\bf P})$ is ``slow and heavy'', whereas
{\bf z} is ``fast and light'':
\begin{equation}
\label{eq:lioutwo}
\epsilon{\partial\phi\over\partial t}+
\epsilon
{{\bf P}\over M}\cdot{\partial\phi\over\partial{\bf R}}-
\epsilon
{\partial h\over\partial{\bf R}}\cdot
{\partial\phi\over\partial{\bf P}}+
\{\phi,h\}=0.
\end{equation}
With this modification, changes in
$({\bf R},{\bf P})$ take place over times of order
unity, whereas changes in {\bf z} occur over times
of order $\epsilon$.

Next, we insert the Ansatz
$\phi=\phi_0+\epsilon\phi_1+\epsilon^2\phi_2+\cdots$
into Eq.\ref{eq:lioutwo}
and order by powers of $\epsilon$:
\begin{eqnarray}
\label{eq:zero}
\Bigl\{h,\phi_0\Bigr\}&=&0\\
\label{eq:r}
\Bigl\{h,\phi_r\Bigr\}&=&
\Biggl({\partial\over\partial t}+
{{\bf P}\over M}\cdot{\partial\over\partial{\bf R}}-
{\partial h\over\partial{\bf R}}\cdot
{\partial\over\partial{\bf P}}\Biggr)
\phi_{r-1},
\end{eqnarray}
$r\ge 1$.
Since $h$ commutes only with functions of itself
under the Poisson bracket (by assumption of ergodicity),
the solution to Eq.\ref{eq:zero} has the form
$\phi_0({\bf R},{\bf P},{\bf z},t)
=A\Bigl({\bf R},{\bf P},h({\bf z},{\bf R}),t\Bigr)$.
To solve for the dependence of $A$ on its arguments,
we must examine Eq.\ref{eq:r}, with $r=1$:
\begin{equation}
\label{eq:one}
\Bigl\{h,\phi_1\Bigr\}=
\Biggl({\partial\over\partial t}+
{{\bf P}\over M}\cdot{\partial\over\partial{\bf R}}-
{\partial h\over\partial{\bf R}}\cdot
{\partial\over\partial{\bf P}}\Biggr)
\phi_0.
\end{equation}
Taking a phase space average of both sides over some
energy shell $(E,{\bf R})$ in the fast phase space,
we get
\begin{equation}
\label{eq:solna}
0={\partial A\over\partial t}+
{{\bf P}\over M}\cdot{\partial A\over\partial{\bf R}}-
{\bf u}\cdot\hat{\bf D}A,
\end{equation}
where the third argument of $A$ is now $E$, the energy
of the shell over which the average is taken.
With the identity
$\hat{\bf D}\cdot(\Sigma{\bf u})=({\bf P}/M)
\cdot\partial\Sigma/\partial{\bf R}$,
we rewrite Eq.\ref{eq:solna} as
\begin{equation}
\label{eq:secondsolna}
{\partial\over\partial t}\Bigl(\Sigma A\Bigr)
=-{{\bf P}\over M}\cdot{\partial\over\partial{\bf R}}
\Bigl(\Sigma A\Bigr)
+\hat{\bf D}\cdot
\Bigl({\bf u}\Sigma A\Bigr).
\end{equation}

To solve for $\phi_1$, we first use Eq.\ref{eq:solna}
to rewrite Eq.\ref{eq:one}:
\begin{equation}
\label{eq:oneagain}
\Bigl\{h,\phi_1\Bigr\}=
-\Biggl({\partial h\over\partial{\bf R}}-{\bf u}
\Biggr)\cdot\hat{\bf D}A\equiv
-{\partial\tilde h\over\partial{\bf R}}
\cdot\hat{\bf D}A,
\end{equation}
adopting the notation of Ref.\cite{br}.
[The left side of this equation is evaluated
at $({\bf R},{\bf P},{\bf z},t)$; the value of the
third argument of $A$ on the right side is
$E=h({\bf z},{\bf R})$.]
This has the form $\{h,f\}=g$; the general solution\cite{br}
consists of both a homogeneous term,
$\phi_{1H}=B({\bf R},{\bf P},h,t)$,
and an inhomogeneous term,
\begin{equation}
\phi_{1I}({\bf R},{\bf P},{\bf z},t)=
\int_{-\infty}^0 ds\,
{\partial\tilde h\over\partial{\bf R}}
({\bf z}_s,{\bf R})\cdot\hat{\bf D}A,
\end{equation}
where ${\bf z}_s({\bf z},{\bf R})$ is the point in phase
space reached by evolving a trajectory from {\bf z}, for a
time $s$, under the Hamiltonian $h({\bf z},{\bf R})$.
Note that $\langle\phi_{1I}\rangle_{E,{\bf R}}=0$
for any $(E,{\bf R})$.

We solve for $B$ much as we did for $A$:
writing Eq.\ref{eq:r}, with $r=2$,
we average each side over an
energy shell $(E,{\bf R})$.
After manipulation, this gives
\begin{equation}
\label{eq:solnb}
0={\partial\over\partial t}\Bigl(\Sigma B\Bigr)+
{{\bf P}\over M}\cdot{\partial\over\partial{\bf R}}
\Bigl(\Sigma B\Bigr)-\hat{\bf D}\cdot
\Bigl({\bf u}\Sigma B\Bigr)-
{1\over 2}
\hat D_i\Bigl(\Sigma L_{ij}\hat D_j A\Bigr),
\end{equation}
where
\begin{equation}
\label{eq:lij}
L_{ij}=2\int_{-\infty}^0 ds
\Biggl\langle{\partial\tilde h\over\partial R_i}
\Biggl({\partial\tilde h\over\partial R_j}\Biggr)_s
\Biggr\rangle_{E,{\bf R}}.
\end{equation}
The first factor inside angular brackets is
evaluated at {\bf z},
the second at ${\bf z}_s({\bf z},{\bf R})$;
the average is over all points {\bf z} on the
energy shell $(E,{\bf R})$.
It is assumed that the integral converges.

Finally, $W({\bf R},{\bf P},E,t)$ is given
by a projection of $\phi$ from
$({\bf R},{\bf P},{\bf z})$ to
$({\bf R},{\bf P},E)$:
\begin{equation}
\label{eq:w}
W=\int d{\bf z}\,\delta(E-h)\phi=
\Sigma(E,{\bf R})\,\langle\phi\rangle_{E,{\bf R}}.
\end{equation}
Since $\langle\phi\rangle_{E,{\bf R}}=A+\epsilon B$
(to order $\epsilon$), we combine
Eqs.\ref{eq:secondsolna}, \ref{eq:solnb}, and \ref{eq:w}
to obtain the desired result, Eq.\ref{eq:central}.

\end{document}